\documentclass[pdflatex]{sn-jnl}
\usepackage{amsmath,amsfonts}
\usepackage{algorithmic}
\usepackage{algorithm}
\usepackage{array}
\usepackage[caption=false,font=normalsize,labelfont=sf,textfont=sf]{subfig}
\usepackage{textcomp}
\usepackage{stfloats}
\usepackage{url}
\usepackage{verbatim}
\usepackage{graphicx}
\usepackage{cite}

\usepackage{color}
\usepackage{varioref,float,amssymb}
\usepackage{amsthm}
\usepackage{booktabs}
\usepackage{hyperref}
\usepackage{colortbl} 
\usepackage{xcolor}   
\usepackage[utf8]{inputenc}
\usepackage[T1]{fontenc}
\usepackage{multirow} 
\usepackage{booktabs}

\newcommand{\ket}[1]{{\left\vert{#1}\right\rangle}}

\usepackage[most]{tcolorbox}

\begin{document}

\title{A Gate-Based Quantum Genetic Algorithm for Real-Valued Global Optimization}

\author[1]{\fnm{Leandro C.} \sur{Souza}}\email{leandro@ci.ufpb.br}

\author[2]{\fnm{Laurent E.} \sur{Dardenne}}\email{dardenne@lncc.br}

\author*[2,3]{\fnm{Renato} \sur{Portugal}}\email{portugal@lncc.br}

\affil[1]{\orgname{Universidade Federal da Paraíba}, \orgaddress{\street{Rua dos Escoteiros, s/n}, \city{Jo\~{a}o Pessoa}, \postcode{58051-900}, \state{PB}, \country{Brazil}}}

\affil[2]{\orgname{National Laboratory of Scientific Computing}, \orgaddress{\street{Av. Getulio Vargas, 333}, \city{Petrópolis}, \postcode{25651-075}, \state{RJ}, \country{Brazil}}}
\affil[3]{\orgname{Universidade Católica de Petrópolis}, \orgaddress{\street{Rua Bar\~{a}o do Amazonas, 124}, \city{Petr\'{o}polis}, \postcode{25685-100}, \state{RJ}, \country{Brazil}}}

\abstract{
We propose a gate-based Quantum Genetic Algorithm (QGA) for real-valued global optimization. In this model, individuals are represented by quantum circuits whose measurement outcomes are decoded into real-valued vectors through binary discretization. Evolutionary operators act directly on circuit structures, allowing mutation and crossover to explore the space of gate-based encodings. Both fixed-depth and variable-depth variants are introduced, enabling either uniform circuit complexity or adaptive structural evolution. Fitness is evaluated through quantum sampling, using the mean decoded output of measurement outcomes as the argument of the objective function. To isolate the impact of quantum resources, we compare gate sets with and without the Hadamard gate, showing that superposition consistently improves convergence and robustness across benchmark functions such as the Rastrigin function. Furthermore, we demonstrate that introducing pairwise inter-individual entanglement in the population accelerates early convergence, revealing that quantum correlations among individuals provide an additional optimization advantage. Together, these results show that both superposition and entanglement enhance the search dynamics of evolutionary quantum algorithms, establishing gate-based QGAs as a promising framework for quantum-enhanced global optimization.

}


\maketitle

\section{Introduction}\label{sec-intro}

Global optimization of real-valued functions is a fundamental challenge in science and engineering, with applications ranging from parameter estimation and machine learning to physics, chemistry, and finance~\cite{horst1995introduction,10.5555/2553119}. Many classical metaheuristic algorithms have been proposed to address this problem, including simulated annealing~\cite{Suman01102006}, particle swarm optimization~\cite{9680690}, differential evolution~\cite{Neri2010}, and genetic algorithms (GAs)~\cite{goldberg1989genetic}. Among these approaches, GAs stand out for their population-based search strategy, which balances exploration and exploitation through stochastic operators such as mutation, crossover, and selection. However, classical GAs often suffer from premature convergence and may struggle in high-dimensional or highly multimodal landscapes~\cite{Katoch2020-wb}.  

Quantum computing~\cite{NC00} offers new opportunities to enhance heuristic optimization methods. By exploiting superposition, entanglement, and interference, quantum circuits can generate probability distributions over candidate solutions that are difficult to reproduce classically. This capability suggests that evolutionary algorithms implemented in the quantum domain could explore solution spaces more efficiently than their classical counterparts. Several quantum-inspired or hybrid proposals have been developed~\cite{Gharehchopogh2023,10.1145/1128022.1128034,computers5040024,Jin_2021,10022159,lahozbeltra2023conquestquantumgeneticalgorithms,10186457,viana2025}, but many of them rely on abstract representations rather than gate-based quantum implementations.  

In this work, we propose a gate-based Quantum Genetic Algorithm (QGA) for real-valued global optimization problems. Each individual in the population is represented by a quantum circuit acting on qubit registers, and its phenotype is obtained by sampling circuit outputs and decoding them into real values. Evolutionary operators act directly on circuit structures, enabling both discrete structural modifications and probabilistic exploration of the solution space. The mutation operator introduces stochastic variations that help the population escape local minima, while the crossover operator promotes the combination and refinement of successful circuit patterns, guiding the search toward the global minimum. We analyze two algorithmic variants: a fixed-depth QGA, in which all circuits share the same number of layers, and a variable-depth QGA, where circuit depth evolves adaptively during evolution, allowing the population to adjust its expressive power dynamically.

To isolate the specific contribution of quantum resources, we compare two configurations that differ only by their gate sets. The \emph{classical gate set} includes only reversible gates that preserve computational basis states, while the \emph{quantum gate set} augments it with the Hadamard gate, enabling coherent superpositions. Simulations on benchmark functions such as Rastrigin, Ackley, and Rosenbrock show that the quantum configuration consistently achieves lower fitness values and faster convergence than the classical one. The improvement correlates with the Shannon entropy of circuit outputs, indicating that quantum state delocalization and superposition enhance exploration of the search space.  

A distinctive contribution of this work is the analysis of \emph{inter-individual entanglement} within the QGA population. In contrast to the previous comparison between classical and quantum gate sets, here we compare two fully quantum configurations that differ only by the presence or absence of entanglement among individuals. By initializing pairs of circuits in maximally entangled states, we show that shared quantum correlations accelerate convergence and lead to systematically lower fitness values than those obtained with independent (non-entangled) quantum populations. This improvement is most pronounced in the early stages of evolution, indicating that entanglement functions as a collective quantum resource that enhances the exploration of the solution space. Together, these results demonstrate that both superposition and entanglement act constructively in the optimization dynamics, providing a genuine quantum advantage within the genetic algorithm framework.

The remainder of this paper is organized as follows. Section~\ref{sec-model} introduces the QGA model, including encoding, initialization, evolutionary operators, and fitness evaluation. Section~\ref{sec:results} presents simulation results, emphasizing the roles of superposition and entanglement in optimization. Finally, Section~\ref{sec:conc} concludes with perspectives on implementation in near-term quantum devices and extensions to higher-dimensional problems.

\section{Quantum Genetic Algorithm Model}\label{sec-model}

Our Quantum Genetic Algorithm (QGA) is designed to find the global minimum of a real-valued function
\[
f : \mathbb{R}^m \longrightarrow \mathbb{R},
\]
where the input is a vector of $m$ real variables \( x_1, x_2, \dots, x_m \). Since quantum circuits operate over discrete quantum states, we discretize each real variable using a binary encoding scheme.

\subsection{Real Variable Encoding}\label{subsec:encode}

We use $n$ qubits to represent each variable \( x_i \), which allows for \( 2^n \) possible binary values per variable. Each $n$-bit string \( b = b_1b_2 \dots b_n \in \{0,1\}^n \) is interpreted as an integer
\[
z = \sum_{j=1}^n b_j \cdot 2^{n-j},
\]
and then mapped to a real number in a user-defined domain \([a_i, b_i] \subset \mathbb{R}\) for variable \( x_i \) via the linear transformation
\[
x_i = a_i + \left( \frac{z}{2^n - 1} \right) (b_i - a_i).
\]
This approach allows for the approximation of both positive and negative real values, depending on the domain \([a_i, b_i]\), and can be made arbitrarily precise by increasing $n$. This encoding serves as the basis for interpreting the outcomes of quantum circuit executions, which we now describe in the context of circuit-based individual representation.

\subsection{Quantum Individuals as Gate-Based Circuits}

In our model, each individual in the population is represented by a quantum circuit that acts on \( m \) quantum registers, each associated with a real-valued variable \( x_i \) and composed of \( n \) qubits. Together, these registers form a single quantum system of \( m n \) qubits. The initial state of the system is a computational basis state \( \ket{b} \in \{0,1\}^{mn} \), selected uniformly at random at initialization. This randomized choice avoids introducing bias into the sampling process and promotes exploration from the outset. The circuit is then built using quantum gates selected from a predefined finite gate set, such as
\[
\mathcal{G} = \{H, S, T, \mathrm{CNOT}\}.
\]

We allow gates acting on at most three qubits in order to keep the model as simple and tractable as possible. Multi-qubit gates, including two- and three-qubit gates, are permitted to act both within a register and across different registers, enabling entanglement between distinct variable representations. The inclusion of three-qubit gates is motivated by classical reversible computation, where universality requires gates acting on at least three bits, such as the Toffoli gate. In our implementation, each three-qubit gate is treated as a single unit and is not decomposed into smaller operations, occupying only one column in the circuit representation. By allowing such gates, our model retains the potential for expressive circuit construction while remaining within a structured and simulable framework.

We consider two variants of the QGA, depending on whether the circuit depth is fixed or allowed to vary during evolution:

\begin{itemize}
    \item In the \textbf{fixed-depth QGA}, each individual has a circuit of fixed depth \( d \), which is treated as a global hyperparameter of the algorithm. The circuit consists of exactly \( d \) layers of gates, and this structure remains constant across the entire population and throughout all generations.

    \item In the \textbf{variable-depth QGA}, each individual may have a different depth \( d \in \mathbb{N} \), which can change over generations. This flexibility allows circuits to grow or shrink as the genetic operators act, enabling dynamic adaptation of circuit complexity during the evolutionary process.
\end{itemize}

In both variants, each gene in the individual's genetic code specifies a gate type from \( \mathcal{G} \) along with its target qubit(s), and the full circuit is defined as a sequence of such gate instructions arranged in layers. The collection of all genes that describe the circuit constitutes the individual's \emph{genotype}, which encodes its structural information. The total number of gates in a circuit is not fixed, but it is upper bounded by \( dnm \), since each of the \( d \) layers can apply at most one gate per qubit. This means that within a single layer, no qubit can be involved in more than one gate, whether it is a one-qubit or a two-qubit operation.

After all gates are applied, the qubits in each of the \( m \) registers are measured in the computational basis. Each register yields an \( n \)-bit string, which represents a discretized value of the corresponding variable \( x_i \). The resulting global bit string of length \( mn \) is thus naturally partitioned into \( m \) substrings of \( n \) bits each, one per register. These substrings are then decoded into approximate real values \( x_1, \dots, x_m \) using the mapping described in Section~3.1. The resulting real-valued vector represents the \emph{phenotype} of the individual (its expressed behavior), on which the fitness evaluation is performed.

The population of the QGA consists of such quantum circuits, each defined by its sequence of gate instructions and depth. The evolutionary process operates on these circuits by applying genetic operators that modify their gate content, structure, and, if permitted, their depth, aiming to produce individuals that sample high-quality solutions to the optimization problem.

\subsection{Circuit Space and Distributional Distance}

To formalize the search domain of the QGA, we define the circuit space \( \mathcal{C}_d(\mathcal{G}) \) as the set of all quantum circuits of fixed depth \( d \) constructed using gates from a finite gate set \( \mathcal{G} \). Each circuit \( U \in \mathcal{C}_d(\mathcal{G}) \) acts on a quantum register of \( mn \) qubits and, upon measurement in the computational basis, induces a probability distribution \( P_U \) over bitstrings of length \( mn \). These bitstrings are then interpreted as real-valued vectors using the encoding described in Section~3.1. Assuming that the gates in \( \mathcal{G} \) have bounded arity, the size of the circuit space satisfies
\[
|\mathcal{C}_d(\mathcal{G})| \in \Theta(2^{mnd}),
\]
where the hidden constant depends on the number and types of gates in \( \mathcal{G} \). This exponential scaling highlights the combinatorial complexity of the genotype space and underscores the importance of effective evolutionary search strategies.

To equip \( \mathcal{C}_d(\mathcal{G}) \) with a geometric structure, we define a distance measure between circuits based on the expected value of their decoded outputs. Each circuit \( C \in \mathcal{C}_d(\mathcal{G}) \) induces a probability distribution \( P_C \) over bitstrings \( z \in \{0,1\}^{mn} \), which are decoded into real-valued vectors \( \mathbf{x}(z) \in \mathbb{R}^m \) using the mapping described in Section~3.1. The expected decoded output of a circuit is given by
\[
\mathbb{E}_{z \sim P_C}[\, \mathbf{x}(z) \,] = \sum_{z \in \{0,1\}^{mn}} P_C(z) \cdot \mathbf{x}(z).
\]
We then define the distance between two circuits \( U, V \in \mathcal{C}_d(\mathcal{G}) \) as the Euclidean distance between their expected decoded outputs:
\[
D_{\mathrm{avg}}(U, V) = \big| \,\mathbb{E}_{z \sim P_U}[\, \mathbf{x}(z) \,] - \mathbb{E}_{z \sim P_V}[\, \mathbf{x}(z) \,] \,\big|.
\]
This metric captures the semantic difference between individuals in the solution space and reflects the actual behavior of the circuits with respect to the optimization task. By operating in the real-valued domain, it naturally incorporates the effect of bit significance and provides a more meaningful notion of distance for evolutionary search.

The space \( \mathcal{C}_d(\mathcal{G}) \), equipped with the average-output metric \( D_{\mathrm{avg}} \), allows us to analyze the behavior of the QGA in geometric terms that reflect the semantic content of the circuits. In this setting, mutation and crossover operators can be interpreted as transformations that move individuals through the space according to changes in their expected decoded outputs. It is important to note, however, that \( D_{\mathrm{avg}} \) defines only a pseudometric on \( \mathcal{C}_d(\mathcal{G}) \), since distinct circuits may induce the same expected decoded output and therefore have zero distance. This occurs, for example, when different circuits generate the same probability distribution over decoded values, despite differing in structure or gate sequence. Besides, distinct circuits may have zero distance if they produce different output distributions that yield the same expected decoded value. As a result, \( D_{\mathrm{avg}} \) captures functional similarity rather than structural identity, which is appropriate in the context of real-valued optimization.

In classical genetic algorithms, mutation typically introduces small, random changes to individual genotypes, supporting local exploration of the search space, while crossover recombines parts of high-performing individuals in the hope of preserving and amplifying beneficial traits. These operators are traditionally seen as complementary: mutation promotes exploration, and crossover facilitates exploitation.

In the quantum setting, however, this distinction becomes more subtle. Since quantum circuits define probability distributions over measurement outcomes, even minor modifications at the circuit level, such as changing a single gate, can produce significant shifts in the output distribution. As a result, the geometric effects of mutation and crossover must be reconsidered in light of the probabilistic nature of quantum circuits. The next sections describe these operators in detail and analyze how they operate within the circuit space \( \mathcal{C}_d(\mathcal{G}) \).


\subsection{Initialization of the Population}

The initial population is defined as a finite sample of circuits drawn from the circuit space \( \mathcal{C}_d(\mathcal{G}) \), which consists of all quantum circuits constructed using gates from the chosen gate set \( \mathcal{G} \). Each individual is generated through a randomized, layer-wise construction process, with sampling rules that depend on both the gate set and the circuit-depth strategy adopted in the experiment.

We consider two types of gate sets in this work:
\begin{itemize}
    \item Classical gate set: \( \mathcal{G}_{\text{cl}} = \{\mathrm{I}, X, \mathrm{CNOT}, \mathrm{SWAP}, \text{Toffoli}, \text{Fredkin}\} \)
    \item Quantum gate set: \( \mathcal{G}_{\text{qu}} = \mathcal{G}_{\text{cl}} \cup \{H, T, T^\dagger, S, S^\dagger, Y, Z\} \)
\end{itemize}

\paragraph{Fixed-depth initialization.}
In the fixed-depth QGA, all circuits in the initial population share the same depth \( d \). We focus on relatively shallow circuits to allow for efficient classical simulation of quantum dynamics. However, shallow circuits tend to produce output distributions with low Shannon entropy, leading to biased sampling over a limited region of the solution space. To mitigate this effect and promote diversity, we adopt a controlled initialization strategy for the first layer when \( \mathcal{G} \) includes the Hadamard gate \( H \), which is the only gate among the considered sets capable of creating superposition from a computational basis state.

For gate sets \( \mathcal{G}_{\text{qu}} \), the first layer of each circuit is constructed by applying Hadamard gates to a randomly selected subset of qubits. Specifically, we draw a random number \( k \in \{1, \dots, mn\} \) and apply Hadamard gates to \( k \) qubits chosen uniformly at random. This randomized application of \( H \) introduces a degree of superposition into the initial population, increasing entropy variation while avoiding uniform or deterministic initialization. The remaining \( d - 1 \) layers are filled by sampling gates uniformly at random from the respective gate set \( \mathcal{G} \), subject to the structural constraint that no qubit is involved in more than one gate per layer. This ensures that the resulting circuits are well-formed and physically realizable.

For the classical gate set \( \mathcal{G}_{\text{cl}} \), no superposition is introduced at any stage. All \( d \) layers of each circuit are constructed entirely at random by sampling gates uniformly from \( \mathcal{G}_{\text{cl}} \), again ensuring that gate placements respect layer-level non-overlapping constraints.

\paragraph{Variable-depth initialization.}
In the variable-depth QGA, each circuit in the initial population is assigned an independent random depth \( d_i \) drawn uniformly from a predefined interval \( [d_{\min}, d_{\max}] \). In this work, we set \( d_{\min} = 1 \) and \( d_{\max} = 10 \), allowing the initial population to contain circuits of varying complexity from very shallow to moderately deep. The construction of each layer follows the same stochastic rules described above, using the selected gate set \( \mathcal{G} \) and enforcing non-overlapping gate placements within each layer. This depth variability introduces an additional source of diversity at the population level, enabling the evolutionary process to explore circuits of different expressiveness from the outset. As evolution proceeds, the depth of each circuit can increase or decrease through genetic operations such as crossover and mutation, allowing the algorithm to adaptively balance expressiveness and efficiency.

This initialization procedure induces a probability distribution over the space of quantum circuits that depends on both the gate set and the chosen depth model. It ensures structural and behavioral diversity at the beginning of the evolutionary process, while enabling or suppressing quantum features such as superposition and entanglement through the appropriate choice of \( \mathcal{G} \).

\subsection{Fitness Evaluation via Quantum Sampling}\label{subsec:fit}

Each individual in the population corresponds to a quantum circuit \( C \in \mathcal{C}_d(\mathcal{G}) \), which defines a probability distribution \( P_C \) over bitstrings in \( \{0,1\}^{mn} \) through measurement in the computational basis. The fitness of an individual is determined by estimating the expected output under this distribution and applying the target function to this estimate.

For each execution (or shot), the circuit \( C \) is run once, and a bitstring \( z^{(s)} \sim P_C \) is sampled. This bitstring is partitioned into \( m \) substrings of \( n \) bits, which are decoded into real-valued variables \( x_1^{(s)}, \dots, x_m^{(s)} \) using the linear mapping described in Section~3.1. After \( N_{\text{shots}} \in \mathbb{N} \) executions, we obtain a sequence of decoded samples \( \{\mathbf{x}^{(s)}\}_{s=1}^{N_{\text{shots}}} \), where \( \mathbf{x}^{(s)} = (x_1^{(s)}, \dots, x_m^{(s)}) \in \mathbb{R}^m \).

The sample mean for each variable is computed as:
\[
x_i^* = \frac{1}{N_{\text{shots}}} \sum_{s=1}^{N_{\text{shots}}} x_i^{(s)}, \quad \text{for } i = 1, \dots, m.
\]
The resulting vector \( \mathbf{x}^* = (x_1^*, \dots, x_m^*) \in \mathbb{R}^m \) serves as an empirical estimator for the expected output:
\[
\mathbf{x}^* \approx \mathbb{E}_{z \sim P_C}[\, \mathbf{x}(z) \,],
\]
where \( \mathbf{x}(z) \) denotes the decoded real vector corresponding to bitstring \( z \).

The fitness of circuit \( C \) is defined by evaluating the objective function at the estimated mean:
\[
\text{fitness}(C) = f(\mathbf{x}^*) = f(x_1^*, \dots, x_m^*),
\]
where \( f : \mathbb{R}^m \to \mathbb{R} \) is the objective function to be minimized.

This sampling-based strategy reflects the statistical nature of quantum computation, where a single circuit defines a distribution over many potential solutions. Unlike classical GAs, which associate each individual with a fixed point in solution space, a quantum individual spans a probabilistic region. Fitness is then interpreted as the function value at the empirical mean, approximating the expected solution encoded in the quantum state.

From a geometric standpoint, the mapping \( C \mapsto \text{fitness}(C) \) defines a real-valued functional on the space \( \mathcal{C}_d(\mathcal{G}) \), which is equipped with the total variation distance between distributions \( P_C \). This induces a metric fitness landscape: circuits that are close in total variation tend to yield similar mean outputs, and hence similar fitness values. This continuity motivates the design of evolutionary operators that act locally within the circuit space.

\subsection{Selection and Reproduction Strategy}

The evolutionary process proceeds in discrete generations, starting from an initial population constructed as described previously. From the second generation onward, each iteration follows a structured sequence of operations combining elitism, crossover, and mutation.

At the beginning of each generation, individuals (represented as circuits \( C \in \mathcal{C}_d(\mathcal{G}) \)) are evaluated and ranked according to their fitness values. A fixed fraction \( p_{\text{elite}} \in (0,1) \) of the population with the best fitness values is selected to form the elite group. These elite individuals are preserved unchanged and directly copied into the next generation, ensuring that high-quality solutions are not lost.

To complete the new generation, the remaining \( 1 - p_{\text{elite}} \) portion of the population is generated via crossover. Pairs of individuals are selected (with or without replacement), and the crossover operator is applied probabilistically to produce offspring. These offspring inherit structural features from their parents, potentially resulting in new circuits that lie in unexplored regions of the circuit space \( \mathcal{C}_d(\mathcal{G}) \).

Following crossover, the mutation operator is applied independently to each offspring, introducing stochastic changes in their gate sequences. Elite individuals are excluded from mutation and remain unmodified. In this formulation, mutation may introduce substantial alterations to a circuit, as it randomly replaces gates and may significantly affect the circuit's output distribution. In contrast, the crossover operator tends to preserve larger subcircuits from each parent, typically producing offspring that remain structurally and functionally closer to their progenitors in the circuit space.

Once crossover and mutation are complete, the full population, composed of preserved elites and newly modified offspring, is re-evaluated, and individuals are re-ranked based on fitness. The next generation then begins.

This strategy maintains a balance between exploitation and exploration in the metric space \( \mathcal{C}_d(\mathcal{G}) \). Elitism exploits the current best regions of the fitness landscape. Crossover explores new combinations across individuals, and mutation enables fine-grained stochastic search. Together, these operations guide the population through the probabilistic landscape defined by quantum circuit outputs.

\subsection{Crossover Operator}

The crossover operator enables structured recombination of quantum circuits within the space \( \mathcal{C}_d(\mathcal{G}) \), allowing the algorithm to explore new regions of the search space by exchanging circuit components between parent individuals. To avoid destructive interference with well-adapted structures, crossover is applied selectively to specific layers.

Crossover is performed with probability \( p_{\text{cross}} \in [0,1] \), a hyperparameter that determines the likelihood of recombination. With probability \( 1 - p_{\text{cross}} \), both parents are copied unchanged into the next generation. When crossover is applied, two parent circuits \( C^{(1)} \) and \( C^{(2)} \) produce two offspring through the exchange of gates in a single layer \( \delta \), chosen at random from the second half of the circuit. This restriction ensures that the earlier layers, which strongly influence the global state evolution, remain stable.

To balance global exploration and local refinement, the amount of information exchanged during crossover is varied. A random integer \( k \in \{1, \dots, n\} \) is drawn, where \( n \) is the number of qubits used to encode each variable. The crossover then interchanges the gates acting on the least significant \( k \) qubits of the selected layer, together with any multi-qubit gates entirely supported within this region. For small \( k \), the operation modifies only the lower part of the register, producing offspring that differ slightly from their parents and thus providing fine-tuning near promising solutions. For larger \( k \), a greater portion of the layer is exchanged, generating more substantial structural changes that enhance population diversity.

This adaptive crossover mechanism naturally produces individuals with varying degrees of similarity to their parents, ensuring that the population contains both exploratory and fine-tuned candidates. In this way, the QGA maintains an effective balance between global search and local optimization throughout the evolutionary process.

\subsection{Mutation Operator}

To maintain genetic diversity in the population and explore new candidate solutions, we introduce a mutation operator that modifies individual circuits with a fixed mutation rate \( p_{\text{mut}} \) per gene. A gene corresponds to a gate instruction at a specific position in the circuit, and mutation is applied independently to each gene with probability \( p_{\text{mut}} \).

The mutation process consists of a gate replacement mechanism. If the replacement is of the same arity (i.e., same number of qubits) as the original, a new gate is drawn uniformly at random from the corresponding subset of \( \mathcal{G} \). If the replacement is of a different arity, additional rules determine how the new gate is placed and how affected qubits are reassigned. The mutation operator handles each case as follows:

\begin{itemize}
    \item \textbf{Case 1: Gate of arity \( k \rightarrow \) Gate of arity \( k \).} The current gate is replaced by another gate of the same arity (where \( k = 1, 2, 3 \)) selected uniformly at random from the corresponding subset of \( \mathcal{G} \), and applied to the same qubit(s).

    \item \textbf{Case 2: One-qubit gate \(\rightarrow\) Two-qubit gate.} The original gate is replaced by a two-qubit gate in which the original qubit becomes the control. The target qubit is selected uniformly at random from among the other qubits at the same layer that currently hold a one-qubit gate. The one-qubit gate on the target is removed. If no such target is available, revert to Case 1.

    \item \textbf{Case 3: One-qubit gate \(\rightarrow\) Three-qubit gate.} The original qubit is used as the control, and two additional target qubits are selected uniformly at random from among the remaining qubits holding one-qubit gates in the same layer. The one-qubit gates on those targets are removed. If fewer than two such qubits are available, revert to Case 1.

    \item \textbf{Case 4: Two-qubit gate \(\rightarrow\) One-qubit gate.} The original two-qubit gate acting on qubits \( q_a \) and \( q_b \) is removed and replaced by two one-qubit gates, each drawn uniformly at random and applied to \( q_a \) and \( q_b \), respectively.

    \item \textbf{Case 5: Three-qubit gate \(\rightarrow\) One-qubit gate.} The original three-qubit gate acting on qubits \( q_a, q_b, q_c \) is removed and replaced by three one-qubit gates, each drawn uniformly at random and applied to \( q_a \), \( q_b \), and \( q_c \), respectively.

    \item \textbf{Case 6: Two-qubit gate \(\rightarrow\) Three-qubit gate.} One of the original qubits is kept as control, and a third qubit is selected uniformly at random from among those holding one-qubit gates in the same layer. That gate is removed, and the new three-qubit gate is applied to the selected triplet. If no such additional qubit is available, revert to Case 1.

    \item \textbf{Case 7: Three-qubit gate \(\rightarrow\) Two-qubit gate.} The three-qubit gate is removed and replaced by a two-qubit gate acting on two of the original three qubits, selected uniformly at random.
\end{itemize}

This mutation mechanism allows for controlled variation in gate types and qubit interactions within each individual circuit, while preserving the layer structure and maintaining the fixed depth \( d \). Reversion to simpler mutations ensures the operator is always well-defined, even under resource constraints at the layer level.

\subsection{Termination Criteria}

The evolutionary process is terminated after a fixed number of generations. Specifically, the algorithm runs for a maximum of \( G_{\text{max}} \in \mathbb{N} \) generations, where \( G_{\text{max}} \) is a user-defined hyperparameter.

This termination criterion ensures a bounded computational cost and provides a consistent stopping rule for all experimental runs. The choice of \( G_{\text{max}} \) balances the trade-off between optimization time and solution quality. In particular, \( G_{\text{max}} \) must be chosen in conjunction with the population size \( |\mathcal{P}| \), as these two hyperparameters jointly determine the total number of individuals evaluated and the overall search effort. A larger population may explore more solutions per generation, while a larger \( G_{\text{max}} \) allows for deeper evolutionary refinement over time.

Table~\ref{tab:hyperparameters} compiles the hyperparameters used in our model. Note that the parameter \( m \), representing the number of real-valued variables in the optimization problem, is not a hyperparameter of the algorithm. It is determined by the dimensionality of the objective function \( f : \mathbb{R}^m \to \mathbb{R} \) and is treated as part of the problem definition.

\begin{table}[!h]
\centering
\begin{tabular}{|c|l|p{8.5cm}|}
\hline
\textbf{Symbol} & \textbf{Name} & \textbf{Description} \\
\hline
$n$ & Qubits per variable & Number of qubits used to encode each real-valued variable $x_i$; determines discretization precision. \\
\hline
$d$ & Circuit depth & Number of gate layers in each quantum circuit (fixed in the fixed-depth QGA). \\
\hline
$p_{\text{mut}}$ & Mutation rate & Probability of mutating each gate instruction (gene) in a circuit. \\
\hline
$p_{\text{cross}}$ & Crossover rate & Probability of applying the crossover operator to a pair of selected parent circuits. \\
\hline
$p_{\text{elite}}$ & Elitism rate & Fraction of top-performing individuals preserved in each generation. \\
\hline
$N_{\text{shots}}$ & Sampling shots & Number of measurements used to evaluate each individual's fitness. \\
\hline
$G_{\text{max}}$ & Max generations & Maximum number of generations for the algorithm to run. \\
\hline
$|\mathcal{P}|$ & Population size & Number of quantum circuits in the population at each generation. \\
\hline
$[a_i, b_i]$ & Variable domain & Real interval for variable $x_i$, used to map bitstrings into real values. \\
\hline
\end{tabular}
\caption{List of hyperparameters in the quantum genetic algorithm}
\label{tab:hyperparameters}
\end{table}



\section{Results}\label{sec:results}

Although the QGA is designed for execution on gate-based quantum hardware, all experiments in this work are conducted through classical simulation of quantum circuits. This choice enables detailed analysis and benchmarking while circumventing current hardware limitations. In particular, we conduct tests both with and without the Hadamard gate \( H \), allowing us to investigate the role of superposition in the optimization process. The simulations, performed with the Qiskit statevector backend, analyze circuits with up to several dozen qubits while preserving the key structural features of the QGA model.

All runs were executed on a high-performance workstation equipped with eight NVIDIA H100 GPUs, using parallel execution to efficiently manage multiple circuit evaluations across independent trials. Each complete run of the genetic algorithm required approximately \textbf{XXX} minutes to finish.

We focus on real-valued global optimization problems using widely studied benchmark functions. In particular, we first consider the Rastrigin function, defined as
\begin{equation}
    f(\mathbf{x}) = 10m + \sum_{i=1}^{m} \left[ x_i^2 - 10 \cos(2\pi x_i) \right],
\end{equation}
where $\mathbf{x} = (x_1, x_2, \ldots, x_m)$ with $x_i \in [-5.12, 5.12]$, and the global minimum is attained at $\mathbf{x} = \mathbf{0}$~\cite{10.1007/BFb0029787}. In this work, we restrict our analysis to functions with two variables. We also consider other two-variable benchmark functions such as Sphere, Ackley, Griewank, and Rosenbrock, which pose diverse challenges including multimodality, narrow valleys, and flat regions. The mathematical expressions of these functions are given in~\cite{Jamil_2013}. The results obtained for these benchmarks are qualitatively similar to those observed for the Rastrigin function, confirming the robustness of the proposed approach across different optimization landscapes. Together, these benchmark functions provide a standard testbed for evaluating the ability of optimization algorithms to escape local minima and effectively explore the global search space.

Throughout all experiments, we fix the following hyperparameters unless otherwise stated: the number of shots used to estimate the output of each quantum circuit is set to \( N_{\text{shots}} = 1024 \); the number of qubits used to encode each real-valued variable is \( n = 8 \), yielding a discretization into \( 2^8 = 256 \) levels; the mutation rate is set to \( p_{\text{mut}} = 30\% \); the crossover rate is set to \( p_{\text{cross}} = 70\% \); and the elitism rate is fixed at \( p_{\text{elite}} = 20\% \). These values were selected to balance exploration and convergence in the QGA dynamics.

The remaining hyperparameters are varied across experiments to assess their impact on performance. In particular, we analyze scenarios where the number of generations \( G_{\max} \), the population size \( |\mathcal{P}| \), and the circuit depth \( d \) are tuned. We typically initialize with \( |\mathcal{P}| = 50 \) and then explore the effects of increasing either  \( G_{\max} \) while keeping $|\mathcal{P}|$ fixed. In the fixed-depth QGA, we investigate the influence of circuit expressiveness by plotting the fitness of the best individual as a function of the circuit depth \( d \). In contrast, for the variable-depth QGA, the depth is not predetermined but evolves as part of the genetic representation, enabling circuits to adapt their complexity dynamically throughout the evolutionary process.

\subsection{Effect of Superposition}\label{subsec:EffectOfSuperposition}

To investigate whether quantum superposition enhances the optimization power of the QGA, we compare two configurations determined by the choice of the gate set \( \mathcal{G} \). In the first configuration, we use the classical gate set: \( \mathcal{G}_{\text{cl}} = \{\mathrm{I}, X, \mathrm{CNOT}, \mathrm{SWAP}, \text{Toffoli}, \text{Fredkin}\} \), whose matrix representations do not generate superposition. These gates preserve computational basis states and thus simulate a classical reversible computation on bitstrings. In the second configuration, we use the quantum gate set: \( \mathcal{G}_{\text{qu}} = \{\mathrm{I}, H, X, Y, Z, T, T^\dagger, S, S^\dagger, \mathrm{CNOT}, \mathrm{SWAP}, \text{Toffoli}, \text{Fredkin}\} \). Since the Hadamard gate \( H \) is also included, this allows circuits to create superposition states and perform non-classical transformations. By comparing the performance of the QGA under these two regimes, we assess the contribution of superposition to the search process and its effect on convergence quality and speed.

\begin{figure}[!h]
    \centering    
    \includegraphics[width=0.6\linewidth]{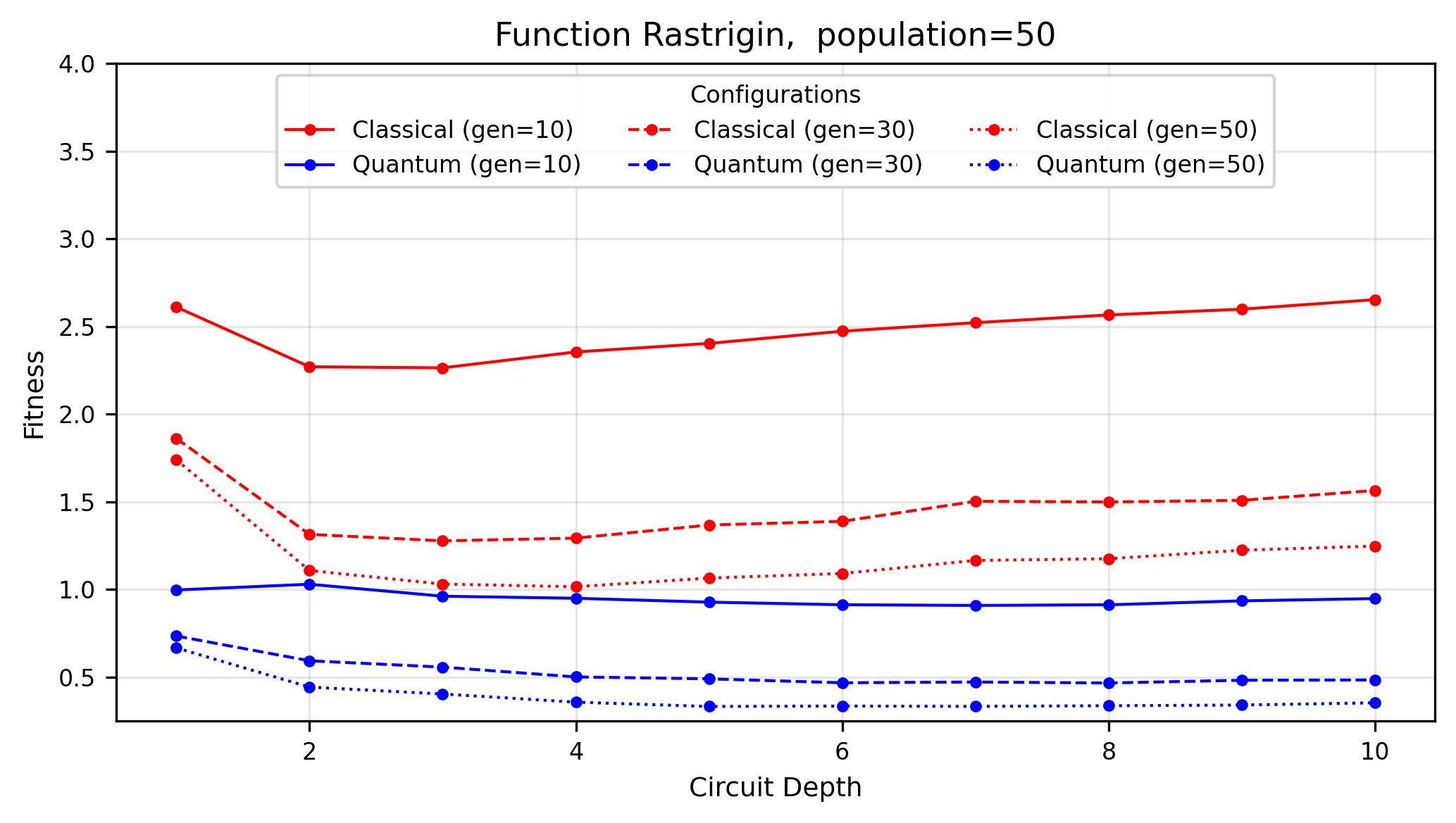}
    \caption{Best fitness obtained on the Rastrigin function as a function of circuit depth for a population size of $|\mathcal{P}|=50$ and different numbers of generations ($G_{\max}=10, 30, 50$). Results are shown for both the classical gate set (red) and the quantum gate set including superposition (blue). Each curve represents the average over 5000 independent repetitions.}
    \label{fig:fit-vs-depth-ras_10_30_50}
\end{figure}

As shown in Figure~\ref{fig:fit-vs-depth-ras_10_30_50}, the inclusion of the Hadamard gate in the quantum gate set consistently improves the optimization performance on the Rastrigin function with two variables ($m=2$) across all generation settings. For $G_{\max}=10$, $30$, and $50$, the quantum configuration (blue curves) achieves lower fitness values than the classical configuration (red curves) at every circuit depth, indicating closer convergence to the global minimum. A clear trend emerges: while classical gates lead to fitness values that gradually deteriorate with increasing depth, the quantum variant maintains robust performance across depths, with only a slight reduction in variance as the number of generations grows. Moreover, by comparing curves for different generation limits, we observe that the best fitness decreases across all depths as $G_{\max}$ increases, confirming that a larger number of generations enhances the overall convergence quality of both classical and quantum configurations. These results confirm that superposition provides a tangible advantage in guiding the evolutionary process, enabling more efficient exploration of the solution space and mitigating the degradation observed in the classical regime.

\begin{figure}[!h]
    \centering
\includegraphics[width=0.6\linewidth]{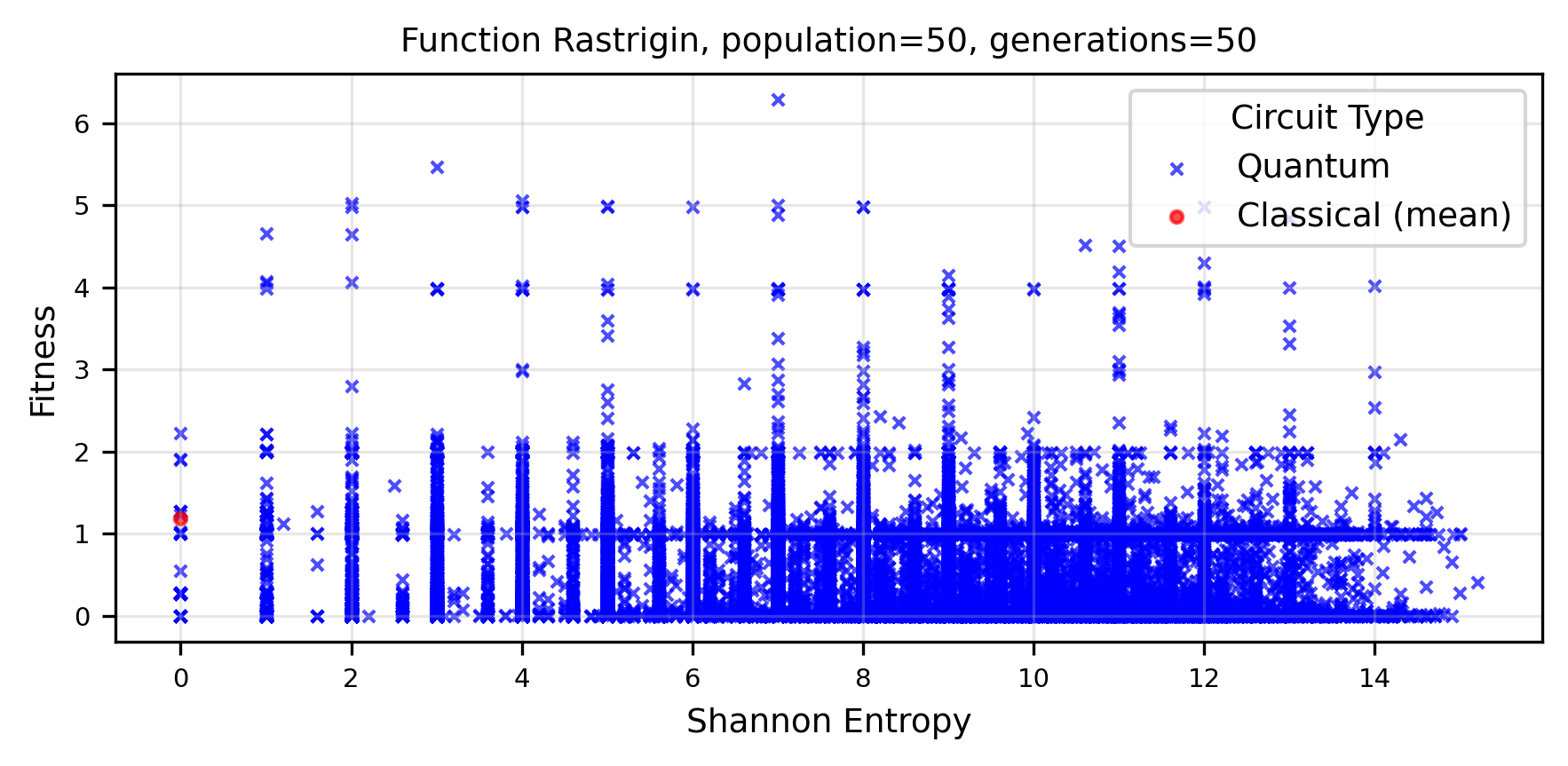}
    \caption{Fitness of the best individuals as a function of the Shannon entropy of the circuits for the Rastrigin function with population size $|\mathcal{P}|=50$ and $G_{\max}=50$. The classical configuration corresponds to the zero-entropy case (red), while the quantum configuration explores a broad entropy spectrum (blue).}
    \label{fig:entropy-ras-50}
\end{figure}

Figure~\ref{fig:entropy-ras-50} illustrates the relation between the Shannon entropy of the quantum circuits and the fitness of the best individuals on the Rastrigin function for $|\mathcal{P}|=50$ and $G_{\max}=50$. Here, the Shannon entropy is computed from the output probability distribution of the circuits in the computational basis, that is, $H(p) = - \sum_x p(x)\,\log p(x)$ with $p(x) = |\langle x|\psi\rangle|^2$, where $\ket{\psi}$ is the output state of the circuit. The classical configuration corresponds to the trivial case of zero entropy, with the mean value obtained after 5000 repetitions, while the quantum configuration explores a wide range of entropy values, with 5000 data points. The results show that low fitness values are consistently associated with higher entropy circuits, indicating that superposition and state delocalization contribute to a more effective exploration of the search space. The equivalent plots for other values of $G_{\max}$ display the same trend, confirming that this behavior is robust to the number of generations. We also note that the entropy values often cluster around integers, a consequence of circuits producing uniform superpositions over $2^k$ basis states, an effect of gates such as $H$ and $T$, whose matrix entries contain factors of $1/\sqrt{2}$, which yield exact entropies of $k$ bits when computed with logarithm base 2.

\subsection{Variable-Depth Circuits}

Having analyzed the role of superposition in fixed-depth circuits, we now turn to the case where the circuit depth is not predetermined but evolves as part of the genetic representation. This variable-depth configuration allows the QGA to adapt the expressiveness of its quantum circuits dynamically during evolution, enabling a balance between search-space exploration and circuit complexity. The analysis presented in this subsection is based on the Rastrigin function, which serves as a representative benchmark; however, qualitatively similar results were obtained for other test functions such as Griewank, Rosenbrock, and Sphere, confirming the generality of the observed behavior. The following results examine how this adaptive mechanism influence

\begin{figure}[!h]
    \centering
\includegraphics[width=0.6\linewidth]{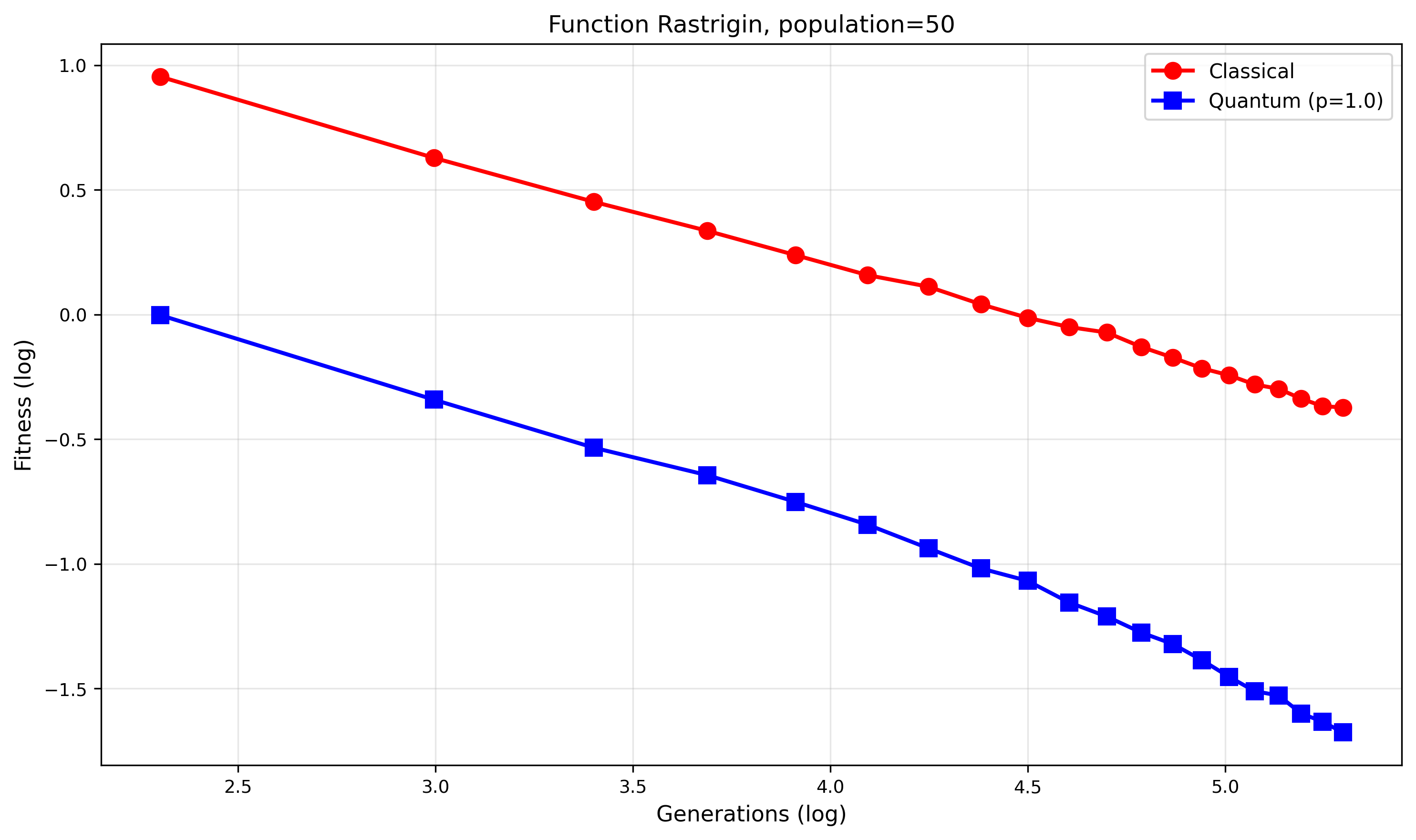}
    \caption{Log–log plot of the fitness of the best individuals as a function of the number of generations for the Rastrigin function with population size $|\mathcal{P}|=50$. Results are shown for both the classical (red) and quantum (blue) configurations, each obtained by averaging over 5000 independent runs.}
    \label{fig:fitness-vs-gen}
\end{figure}

Figure~\ref{fig:fitness-vs-gen} shows the evolution of the fitness of the best individuals as a function of the number of generations for the Rastrigin function with population size $|\mathcal{P}|=50$. Both the classical and quantum configurations display a monotonic decrease in fitness with increasing generations, consistent with progressive convergence toward the global minimum. In the log–log scale, the curves exhibit an approximately linear trend, suggesting a power-law relationship between fitness improvement and the number of generations. The quantum configuration (blue) lies systematically below the classical one (red) and presents a steeper slope, indicating faster convergence and more effective exploration of the search space. Disregarding the last few points, the extrapolated behavior suggests that the algorithm approaches the global minimum within roughly ten generations, after which improvements become marginal. This plot was obtained by averaging the results over 5000 independent runs. 

\begin{figure}[!h]
    \centering
\includegraphics[width=0.6\linewidth]{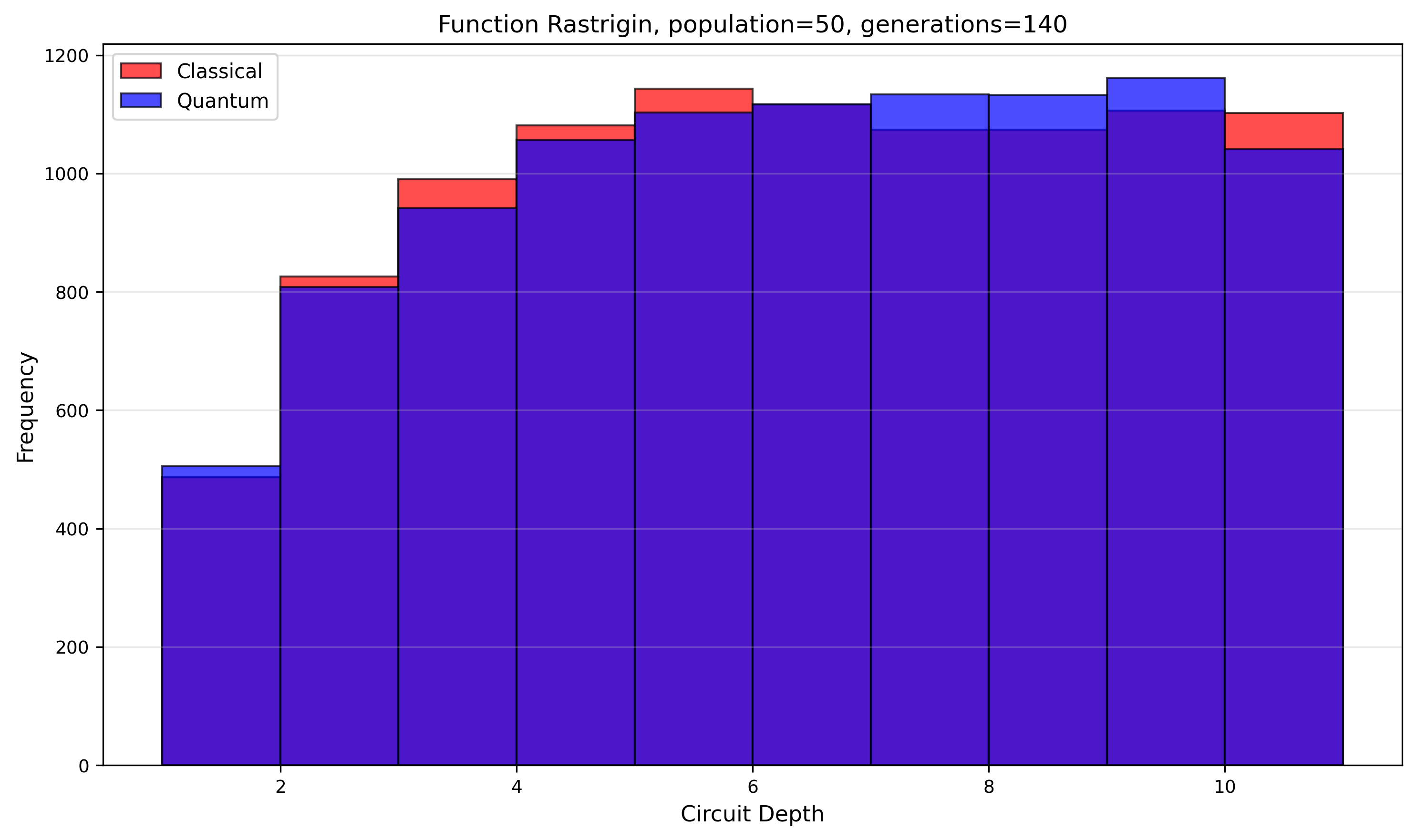}
   \caption{Circuit depth distribution after 140 generations for the Rastrigin function with population size $|\mathcal{P}|=50$. Both the classical (red) and quantum (blue) configurations favor deeper circuits, reflecting the evolutionary tendency toward greater expressiveness. Results are averaged over 5000 runs.}
    \label{fig:depth-histogram}
\end{figure}

Figure~\ref{fig:depth-histogram} shows the distribution of circuit depths after 140 generations for the Rastrigin function with population size $|\mathcal{P}|=50$. Both the classical and quantum configurations exhibit a clear tendency toward deeper circuits, indicating that the evolutionary process favors individuals with greater expressive power. This observation aligns with the results discussed in Section~\ref{subsec:EffectOfSuperposition}, where quantum circuits incorporating superposition achieved lower fitness values across all depths. Here, the variable-depth mechanism reinforces that trend: as the algorithm evolves, it naturally promotes deeper and more expressive circuits, while the inclusion of quantum gates such as $H$ and $T$ further enhances their representational capacity. The similar overall shape of the two distributions demonstrates that both configurations exploit circuit complexity, but the quantum version leverages superposition to extract greater optimization gains from comparable structural depth.

\begin{figure}[!h]
    \centering
\includegraphics[width=0.6\linewidth]{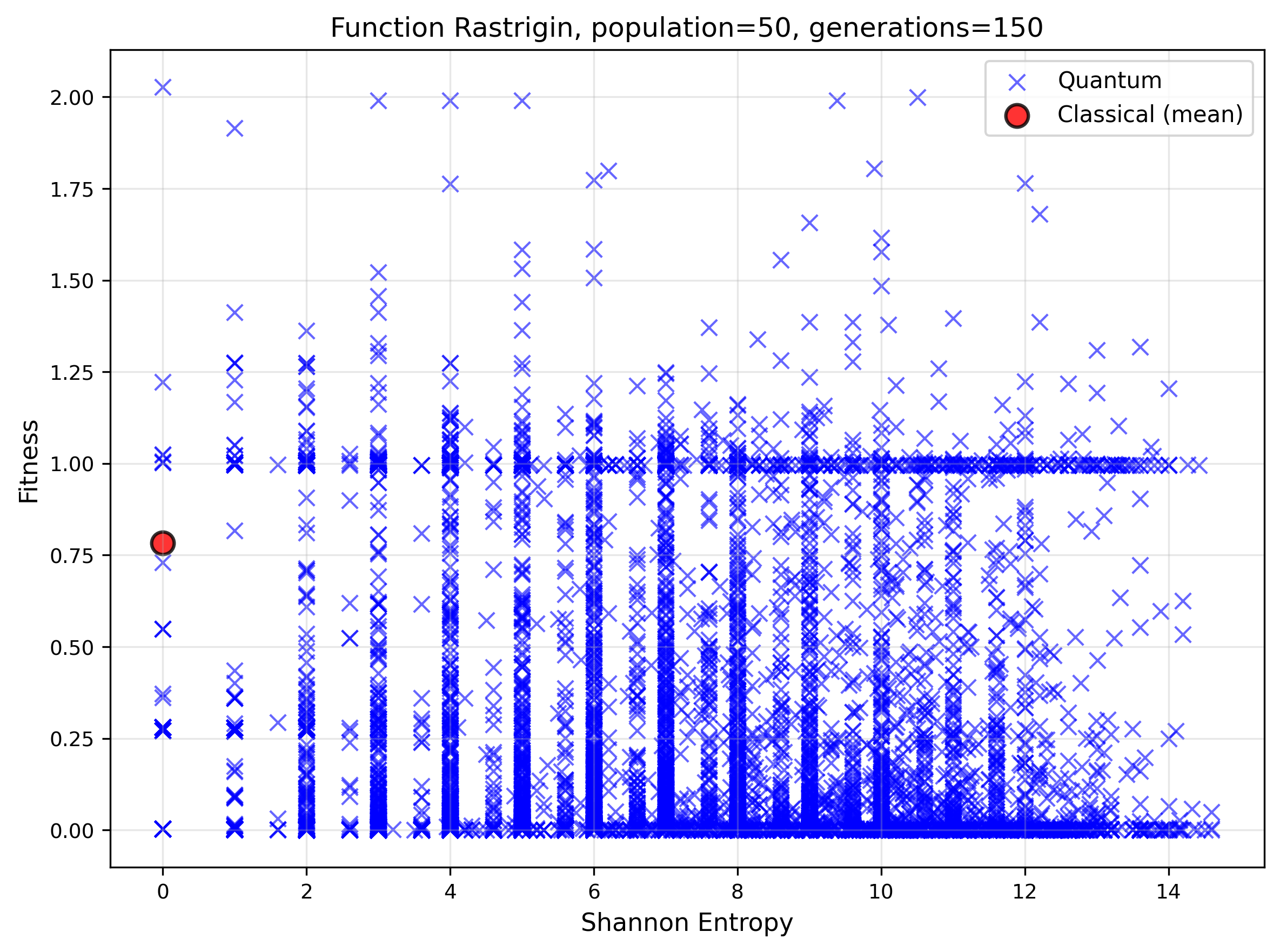}
  \caption{Fitness versus Shannon entropy after 150 generations for the Rastrigin function with population size $|\mathcal{P}|=50$. The classical configuration (red) remains at zero entropy, while the quantum configuration (blue) spans a broad range of values, with higher entropy correlating with improved fitness. Results are averaged over 5000 runs.}
    \label{fig:fitness-vs-shannon}
\end{figure}

Figure~\ref{fig:fitness-vs-shannon} shows the plot of fitness versus Shannon entropy for the circuits obtained after 150 generations on the Rastrigin function with population size $|\mathcal{P}|=50$. The classical configuration (red) corresponds to the trivial case of zero entropy, while the quantum configuration (blue) spans a wide range of entropy values. As observed in Section~\ref{subsec:EffectOfSuperposition}, higher-entropy circuits are generally associated with lower fitness values, confirming that quantum superposition and state delocalization contribute to a more efficient search process. The clustering of points around integer entropy values was discussed in connection with Figure~\ref{fig:entropy-ras-50}, while the concentration of points at fitness equal to one arises from the fact that the Rastrigin function contains many local minima with this value and none with fitness greater than zero but smaller than one.

\subsection{Effect of Entanglement}

To investigate how entanglement among individuals affects the optimization process, we compare two populations that evolve under identical conditions using the quantum gate set \( \mathcal{G}_{\text{qu}} \). In both cases, each individual circuit may internally generate entanglement during its evolution, as described previously. The distinction lies in the presence or absence of entanglement between individuals in the initial setup of the population. In the first configuration, the population is initialized with no inter-individual entanglement: all individuals start independently, and their quantum states remain separable throughout the entire evolutionary process. In the second configuration, the population is initialized with pairwise maximal entanglement, where individuals are grouped into randomly chosen pairs that share maximally entangled states at the beginning of the evolution. This entanglement pattern is applied consistently across generations, while the genetic algorithm proceeds as usual with the same selection, crossover, and mutation operators. This setup enables the evaluation of how inter-individual entanglement influences the convergence behavior and whether shared quantum correlations among individuals can enhance the overall optimization performance of the QGA.

\begin{figure}[!h]
    \centering
    \includegraphics[width=0.6\linewidth]{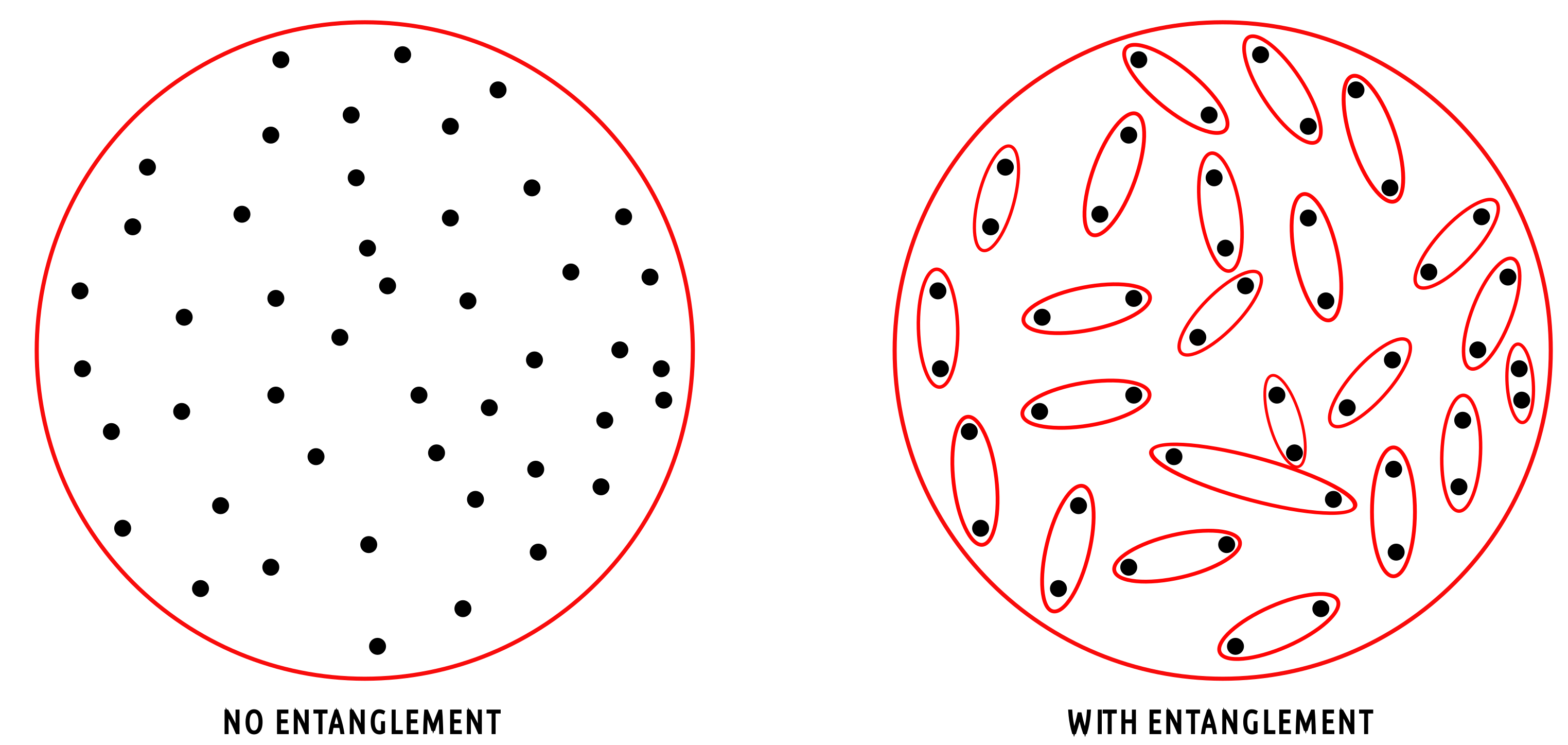}
    \caption{Schematic illustration of populations without (left) and with (right) pairwise inter-individual entanglement. Each red ellipse represents a randomly chosen pair of individuals prepared in a maximally entangled state.}
    \label{fig:entangle-vs-nonentangle}
\end{figure}

Note that in this case we compare two versions of the Quantum Genetic Algorithm: one with pairwise inter-individual entanglement and another with a population of non-entangled individuals, as illustrated in Figure~\ref{fig:entangle-vs-nonentangle}. We restrict our analysis to pairwise entanglement because each simulation must handle two individuals (two circuits) simultaneously, requiring 16 qubits for one-variable functions and 32 qubits for two-variable functions.

\begin{figure}[!h]
    \centering
    \includegraphics[width=0.6\linewidth]{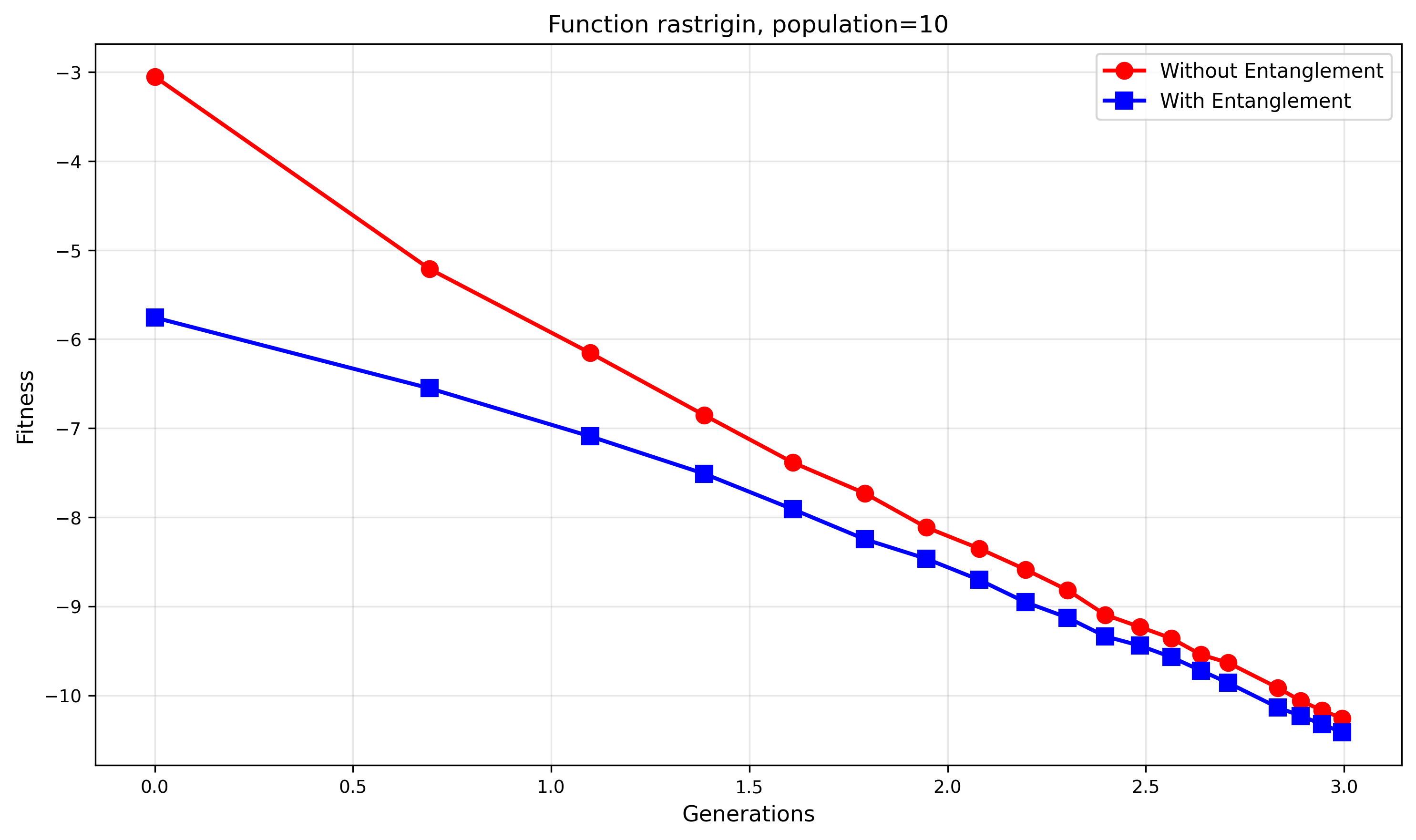}
    \caption{Fitness versus number of generations for the Rastrigin function with population size $|\mathcal{P}|=10$, comparing populations without inter-individual entanglement (red) and with pairwise entanglement (blue). The entangled population consistently attains lower fitness values, indicating faster convergence. Results are averaged over 5000 runs.}
    \label{fig:fitness-vs-entanglement-pop10}
\end{figure}

Figure~\ref{fig:fitness-vs-entanglement-pop10} compares the evolution of the fitness as a function of the number of generations for populations with and without inter-individual entanglement on the Rastrigin function with $|\mathcal{P}| = 10$. Both populations display monotonic improvement in fitness, confirming consistent convergence of the QGA. However, the population initialized with pairwise entanglement (blue) systematically achieves lower fitness values than the non-entangled population (red) across all generations. This result indicates that shared quantum correlations among individuals enhance the collective search capability of the algorithm, allowing faster convergence toward high-quality solutions. The performance gap is most evident during the early stages of evolution, suggesting that initial entanglement improves exploratory efficiency before both populations approach similar asymptotic fitness values.

\section{Conclusion}\label{sec:conc}

In this work, we introduced a gate-based Quantum Genetic Algorithm (QGA) for real-valued global optimization, where individuals are explicitly represented as quantum circuits whose measurement statistics encode candidate solutions. Evolutionary operators act directly on circuit structures, enabling both probabilistic and structural exploration of the search space within a unified quantum–evolutionary framework. Each circuit defines a probability distribution over real vectors, enabling the algorithm to leverage intrinsic quantum resources, such as superposition and entanglement. The model also incorporates a circuit-space pseudometric that relates genotypic variation to functional behavior, providing a geometric interpretation of mutation and crossover as transformations in a probabilistic pseudometric space. Within this space, mutation typically produces large stochastic jumps that promote exploration of distant regions, while crossover can generate both substantial transitions and fine-grained refinements, guiding the population toward the global minimum.

Extensive simulations on benchmark functions, including Rastrigin, Ackley, Griewank, Sphere, and Rosenbrock, show that the use of quantum gate sets containing the Hadamard gate consistently enhances convergence speed and final solution quality relative to classical reversible configurations. Circuits exhibiting higher Shannon entropy, which are indicative of richer superposition, tend to achieve lower fitness values, confirming that quantum state delocalization facilitates global exploration. Moreover, the variable-depth version of the QGA enables adaptive adjustment of circuit expressiveness, balancing optimization accuracy and computational cost across generations.

A distinct and central finding of this study concerns the role of inter-individual entanglement. When individuals in the initial population are arranged as randomly chosen, pairwise maximally entangled states, the QGA exhibits systematically faster convergence and achieves lower fitness values than when individuals evolve independently. This improvement occurs even though both configurations employ the same quantum gate set, demonstrating that entanglement between individuals constitutes an additional quantum resource, distinct from superposition, that enhances collective search dynamics and accelerates early-stage exploration of the fitness landscape.

Taken together, these results establish that both intra-circuit quantum effects (superposition within individuals) and inter-circuit quantum correlations (entanglement across individuals) contribute constructively to optimization performance. The proposed gate-based QGA provides a scalable framework for exploring these phenomena on near-term quantum devices and motivates future extensions involving adaptive entanglement topologies, multi-individual correlations, and advanced theoretical modeling of circuit-space dynamics.

Future research directions include implementing the proposed QGA on near-term quantum hardware to evaluate its performance under realistic noise and decoherence, as well as extending the model to higher-dimensional and constrained optimization problems. Another promising avenue is the development of adaptive entanglement strategies, allowing the degree and topology of inter-individual entanglement to evolve dynamically throughout the optimization process. Exploring multi-individual entanglement networks beyond pairwise coupling could reveal new forms of collective quantum behavior and further enhance search efficiency.

Additional investigations may focus on alternative fitness evaluation schemes that incorporate statistical properties of the sampled distributions, and on analytical modeling of convergence using geometric and information-theoretic tools to relate circuit-space structure, entropy, and optimization dynamics. Finally, systematic comparisons with other classes of quantum evolutionary algorithms, such as quantum particle swarm and quantum differential evolution, could help clarify the unique advantages of gate-based QGAs in practical quantum optimization.

\section*{Acknowledgements}

We thank Pedro Portugal for insightful discussions. The work of L.~C.~Souza was supported by CNPq grant number 302519/2024-6. The work of R.~Portugal was supported by FAPERJ grant number CNE E-26/200.954/2022, and CNPq grant numbers 304645/2023-0 and 409552/2022-4.



\end{document}